\pdfoutput=1

\documentclass[%
twocolumn,
reprint,
superscriptaddress,
preprintnumbers,
nofootinbib,
showpacs,
aps,
prl
]{revtex4-2}

\usepackage[utf8]{inputenc}
\usepackage{graphicx, color}
\usepackage[svgnames,dvipsnames]{xcolor}
\usepackage{ragged2e}
\usepackage{dcolumn}
\usepackage{bm}
\usepackage{amsmath}
\usepackage{amsthm}
\usepackage{ascmac}
\usepackage{xparse} 
\usepackage{mathtools} 
\usepackage{blkarray} 
\usepackage{amscd,verbatim}
\usepackage{amssymb}
\usepackage[all]{xy}
\usepackage{tikz}
\usepackage{tikz-cd}
\usepackage{pgf,pgfplots}
\usepackage{enumitem}
\usepackage{slashed}
\usepackage{comment}
\usepackage{footmisc}
\usepackage{braket}
\usepackage{listofitems} 
\usetikzlibrary{arrows.meta} 
\usepackage[outline]{contour} 
\usepackage[colorlinks,citecolor=DarkGreen,linkcolor=FireBrick,linktoc=all]{hyperref}
\pgfplotsset{compat=newest}

\contourlength{1.4pt}

\usetikzlibrary{decorations.pathmorphing}

\tikzset{>=latex} 

\newcommand{\ssection}[1]{
\vspace{6pt}
\textit{#1}. --- }
\renewcommand{\d}[0]{\mathrm{d}}

\newcommand{\red}[1]{{#1}}

\allowdisplaybreaks

\begin{document}
\preprint{preprint}

\title{
Lindblad dynamics in holography
}

\author{Takanori Ishii}
\email{ishiit@gauge.scphys.kyoto-u.ac.jp}
\affiliation{Theoretical Particle Physics Group
Department of Physics, Kyoto University
Kitashirakawa, Kyoto 606-8502, Japan}
\preprint{KUNS-3047}

\author{Daichi Takeda}
\email{daichi.takeda@riken.jp\\ \textcolor{black}{The authors equally contributed.}}
\affiliation{iTHEMS, RIKEN, Wako, Saitama 351-0198, Japan}
\preprint{RIKEN-iTHEMS-Report-25}

\begin{abstract}
We develop, in the AdS/CFT correspondence, a method to compute correlation functions when the CFT is governed by the Lindblad equation for open quantum systems, via the AdS theory.
Using a simple example in AdS$_3$/CFT$_2$, we demonstrate that the predictions of the AdS theory based on our method match the direct computations in the dual CFT.
\red{We also briefly discuss the relaxation problem and the holographic entropy in this example.}
\end{abstract}

\maketitle

\ssection{Introduction}
The AdS/CFT correspondence \cite{Maldacena:1997re, Gubser:1998bc, Witten:1998qj}, embodying the holographic principle, conjectures an equivalence between $(d+1)$-dimensional gravity in anti–de Sitter (AdS) spacetime (the “bulk”) and $d$-dimensional conformal field theories (CFT, the “boundary”).  
AdS/CFT has enabled probing strongly coupled quantum field theories (QFT), including condensed-matter systems, via bulk perturbative calculation.  
Conversely, boundary CFTs provide insights into the microscopic structure of gravity.

In particular, AdS/CFT can address thermal physics using the holographic Schwinger-Keldysh (SK) formalism \cite{Son:2002sd, Herzog:2002pc, Skenderis:2008dh, Skenderis:2008dg, vanRees:2009rw}.
Ref \cite{Skenderis:2008dh, Skenderis:2008dg, vanRees:2009rw} finally established a way to build bulk spacetimes dual to SK-like contours, where we glue several Lorentzian and Euclidean spacetimes.
This enables us to predict various correlators of out-of-equilibrium quantum systems in the bulk terms.
A simplified method was also proposed in \cite{Glorioso:2018mmw}.

Not limited to that, AdS/CFT is also applicable to open systems.
For example, \cite{Son:2009vu, deBoer:2008gu} initiated the study of the Brownian motion of a test quark in the holographic thermal bath, using the bulk Nambu–Goto string in a black brane background.
Similarly, \cite{Jana:2020vyx} developed a method to derive effective theories of open QFT, where they couple the QFT to a holographic bath and trace-out the bath's degrees of freedom by solving the bulk EOM.
Conversely, in the scenario in which the holographic CFTs themselves are open and interact with each other, the the Higgs mechanism \cite{Aharony:2006hz, Karch:2023wui} and the change of the Hilbert space \cite{Geng:2023ynk} were investigated.

Despite those remarkable progress, AdS/CFT still cannot handle the Lindblad equation \cite{Gorini:1975nb, Lindblad:1975ef}, the most fundamental dynamics of open quantum systems; the corresponding bulk picture remains unknown.
Because many real-world systems --- from condensed matter to black holes --- frequently undergo non-unitary evolution, the holographic Lindblad dynamics is essential for AdS/CFT to be more realistic.

The purpose of this Letter is to provide an answer.
We consider the boundary CFT under the Lindblad equation and derive the dual bulk description that provides boundary correlators; the boundary theory is not a bath, but rather the target open system.
As in the SK formalism, the bulk spacetime consists of forward and backward Lorentzian segments, with white noise entering the asymptotic boundary condition --- this noise reflects the non-unitarity.
To verify the method, we apply it to a real free scalar field on AdS$_3$ and explicitly compute the one- and two-point functions.
We confirm our model matches the Lindblad dynamics of a generalized free CFT$_2$ with the initial state chosen as the vacuum.

Our framework will pave a way for exploring, via AdS/CFT, open quantum systems and even open gravitational systems.
For instance, our method can incorporate environmental effects into AdS/CMP \cite{Hartnoll:2009sz, Herzog:2009xv,McGreevy:2009xe, Horowitz:2010gk, Sachdev:2010ch}, which studies condensed matter phenomena via AdS/CFT.
Moreover, in the context of black hole evaporation, we expect our framework to provide an alternative prescription for letting Hawking radiation escape from AdS, similar to approaches that explicitly attach baths \cite{Rocha:2008fe, Almheiri:2019psf, Penington:2019npb, Almheiri:2019qdq}\red{. Also, \cite{Glatthard:2025mbb} pointed out that some Lindblad systems share analogous entropy evolution with black hole evaporation processes.}

\ssection{Bulk description of boundary Lindblad dynamics}
We present a method to compute, in the gravity language, the generating functional that yields the boundary correlators of Lindblad dynamics.

\begin{figure}[b]
    \centering
    \includegraphics[scale = 1]{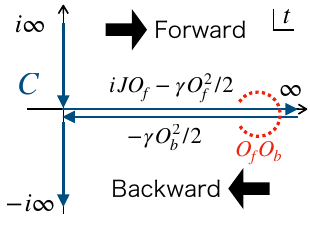}
    \caption{
    The boundary spacetime $C$ in \eqref{eq: boundary Z}.
    Only the time contour is drawn here.
    The imaginary time evolutions are just for making the initial state.
    }
    \label{fig: boundary_contour}
\end{figure}

We begin with a generic $d$-dimensional holographic CFT.
Let $H$ be the Hamiltonian, and $O(\vec x)$ a primary operator in the Schr\"odinger picture, where $\vec x$ denotes the $(d-1)$-dimensional spacelike coordinates.
For simplicity, we further assume that $O$ is real, spin-$0$, and independent of the conjugate momenta of elemental fields.
We consider the following Lindblad equation:
\begin{align}\label{eq: Lindblad}
  \dot \rho(t) =& \mathcal{L} \rho(t),\\
  \mathcal{L} A :=& -i[H, A]\nonumber\\
    &+ \gamma \int \d^{d-1}\vec x\,\left[O(\vec x)A O(\vec x) - \frac{1}{2}\{O(\vec x)^2,A\}\right].
\end{align}
Here, the jump operator is chosen as $O$, and $\gamma$ is a positive constant (the extension to position-dependent $\gamma$ is straightforward).
The equation can be solved formally as $\rho(t) = e^{t \mathcal{L}}\rho(0)$.

The time-ordered correlators for Markovian systems \cite{gullo2014non} are given, for $0<t_1<\cdots <t_p$, as
\begin{align}\label{eq: correlators}
    &\braket{O(t_p,\vec x_p)\cdots O(t_1,\vec x_1)} :=\nonumber\\
    &\mathrm{Tr}\left[
    O(\vec x_p)e^{(t_p-t_{p-1})\mathcal{L}} O(\vec x_{p-1}) \cdots O(\vec x_1) e^{t_1\mathcal{L}} \rho(0)
    \right],
\end{align}
where each superoperator acts on the entire expression of its right.
While we focus on correlators of the jump operator $O$, our framework below works for other correlators with minor modifications.

The generating functional to yield \eqref{eq: correlators} is given \cite{WalterTStrunz_1997, Meng:2020czp} by
\begin{widetext}
    \begin{align}\label{eq: boundary Z}
        Z[J] := \int\mathcal{D}\phi\, e^{iI[C;\phi]} \exp\left[\int_{t=0}^{t=\infty} \d^{d}x\,\left(i J(x) O_f(x) + \gamma O_f(x) O_b(x)- \frac{\gamma }{2}O_f(x)^2 - \frac{\gamma }{2}O_b(x)^2\right)\right],
    \end{align}
\end{widetext}
where $\phi$ collectively denotes the elementary fields of the CFT, $I$ is the CFT action, $C$ is the glued spacetime, and the subscripts $f$ and $b$ refer to the forward and backward segments respectively, with $x = (t,\vec x)$.
FIG.\ \ref{fig: boundary_contour} illustrates the path integral contour.
The two vertical segments in FIG.\ \ref{fig: boundary_contour} represent the Euclidean time evolution that prepares the initial state $\rho(0)$.
Generally, initial states whose AdS/CFT interpretation is clear are those prepared by Euclidean path integrals, so we consider that $\rho(0)$ is also prepared in that manner.
For mixed $\rho(0)$, one can purify it in advance.

Let us move onto the bulk picture.
In the standard AdS/CFT ($\gamma = 0$), the duality is expressed by the equivalence between the generating functionals of both theories \cite{Maldacena:1997re, Gubser:1998bc, Witten:1998qj}.
The bulk spacetime metric is assumed to be asymptotically AdS, meaning that in Lorentzian signature it behaves as
\begin{align}\label{eq: asymptotic AdS}
    \d s^2 \sim \frac{\d r^2}{r^2} + r^2(-\d t^2 + \d \vec x^2)
    \quad
    (\mbox{as $r\to \infty$}),
\end{align}
where $r$ is the radial coordinate corresponding to the extra dimension.
In most cases, we are interested in flat, spherical, or hyperbolic space for the spacelike directions $\vec x$.
Another essential bulk field is the real scalar field $\Phi$ dual to $O$.
It satisfies the asymptotic boundary condition $\Phi \sim r^{\Delta - d}\,J$ as $r \to \infty$, where $J$ is the CFT external source in \eqref{eq: boundary Z} and $\Delta$ is the scaling dimension of $O$, by which the mass of $\Phi$ is given as $m^2 = \Delta(\Delta-2)$.

\begin{figure}
    \centering
    \includegraphics[scale = 1]{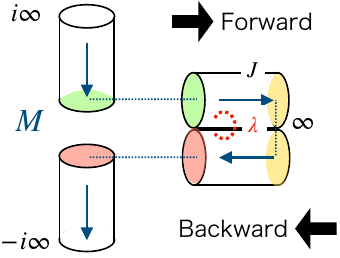}
    \caption{The bulk spacetime $M$ to correspond to $C$ in FIG. \ref{fig: boundary_contour}.
    The fields are identified at the junction surfaces of the same color.
    $J$ and $\lambda$ are in the boundary condition \eqref{eq: BC}.
    }
    \label{fig: bulk_contour}
\end{figure}

To obtain the bulk picture for $\gamma \neq 0$, we need to deform \eqref{eq: boundary Z} so that the standard AdS/CFT duality applies.
Such a technique has already been developed in \cite{Aharony:2001pa, Witten:2001ua, Berkooz:2002ug, Sever:2002fk, Aharony:2005sh, Aharony:2006hz}, and we only need to introduce auxiliary fields.
Following that strategy, we can rewrite $Z[J]$ in terms of the bulk gravity \cite{supp} as
\begin{align}\label{eq: bulk Z}
    Z[J] &= \int \mathbb{D}\Phi \mathcal{D}\lambda\; e^{i S[M;\Phi;J + \lambda, \lambda]- \frac{1}{2\gamma}\int \d^d x\; \lambda(x)^2}.
\end{align}
Above, $S$ is the bulk action, $\mathbb{D}$ is the bulk path integral measure, $M$ is the spacetime manifold whose asymptotic boundary constitutes $C$ (FIG.\ \ref{fig: bulk_contour}), and the argument $[\cdots;J + \lambda, \lambda]$ in $S$ indicates that $\Phi$ is subject to the following boundary condition:
\begin{align}\label{eq: BC}
    \Phi_f \sim r^{\Delta-d} (J + \lambda),\quad
    \Phi_b \sim r^{\Delta-d} \lambda\quad
    (\mbox{as $r\to \infty$}).
\end{align}
Similar boundary conditions are imposed on the Euclidean segments, depending on the boundary initial state.
In \eqref{eq: bulk Z}, the time domain of $\d^{d}x$ is $[0, \infty)$, and all fields take the same value on each junction surface (FIG.\ \ref{fig: bulk_contour}).
For notational simplicity, we have omitted bulk fields other than $\Phi$, but implicitly included them in the path integral measure.

We usually consider the large $N$ (or large central charge) limit, in which the classical approximation holds in the bulk.
Taking the limit while fixing $\mathcal{\gamma}$, we can evaluate the bulk path integral by the on-shell action, after which we perform the $\lambda$-integral.

Here is a remark before turning to the example.
Ref \cite{chenu2017quantum} reformulated the Lindblad equation as a unitary time evolution with white noise.
From this perspective, one can more easily derive \eqref{eq: bulk Z} by applying the conventional real-time AdS/CFT \cite{Skenderis:2008dg, Skenderis:2008dh}, where the $\lambda$-Gaussian factor in \eqref{eq: bulk Z} is nothing but the ensemble average of the white noise.
\red{However, one should note that the formalism of \cite{chenu2017quantum} does not cover the general class of Lindblad equations, as it requires the total Hamiltonian including white noise term to be Hermitian, which is trivially satisfied in the current case where the jump operators $O(\vec x)$ are all real.}

\ssection{Example: bulk free scalar theory}
We here consider $\Phi$ to be a free field with the background fixed to AdS$_3$, and derive the one- and two-point functions of its dual CFT.

The model is given as
\begin{align}\label{eq: bulk action}
    &S = -\frac{1}{2}\int \d^3 x \sqrt{|g|} \left[g^{\mu\nu}\partial_\mu \Phi \partial_\nu \Phi +m^2\Phi^2\right] + S_\mathrm{ct},\\
    &g_{\mu\nu}\d x^\mu \d x^\nu = -(1+r^2)\d t^2 +\frac{\d r^2}{1+r^2} +r^2 \d\theta^2,
\end{align}
where $S_\mathrm{ct}$ is the counterterm \cite{deHaro:2000vlm} and $\theta$ is $2\pi$-periodic.
In the Euclidean segments, the line element $\d t$ becomes purely-imaginary.

Let us evaluate $Z[J]$ of \eqref{eq: bulk Z} with \eqref{eq: bulk action}, in the large $N$ limit.
First, we fix $\lambda$ and find the classical solution subject to \eqref{eq: BC}.
For the Euclidean segments, we impose
\begin{align}\label{eq: Euclidean BC}
    \Phi_\mathrm{E} \sim 0 \times r^{\Delta-d}
    \qquad
    (r\to \infty),
\end{align}
due to $\rho(0) = \ket 0\hspace{-3pt} \bra 0$.
The junction conditions on each gluing surface in FIG.\ \ref{fig: bulk_contour} are provided in \cite{Skenderis:2008dg, Skenderis:2008dh}.
Then, we evaluate \eqref{eq: bulk action} with the solution and perform the remaining $\lambda$-integral with the on-shell action.
In this case, the integral becomes a Gaussian integral, and can be performed completely.

Ultimately, we obtain the generating functional \cite{supp} as
\begin{align}\label{eq: example Z}
    Z[J] &= \exp\left(S_\mathrm{uni}[J] + S_\mathrm{diss}[J] \right),
\end{align}
where we define
\begin{widetext}
\begin{align}
    S_\mathrm{uni}[J] &:= -\frac{2(\Delta-1)^2}{\pi} \int \d t \d\theta \d\hat{t}\d\hat{\theta}\; \Theta(t-\hat{t}) F_{\Delta}(t-\hat{t},\theta-\hat{\theta})J(t,\theta)J(\hat{t},\hat{\theta}),\label{eq: S_uni}
\\
    S_\mathrm{diss}[J] &:= \frac{2\gamma(\Delta-1)^4}{\pi^2} \int \d t\d\theta \d\hat{t}\d\hat{\theta} \d t^\prime \d\theta^\prime\;\Theta(t-t^\prime)\Theta(\hat{t}-t^\prime) J(t,\theta)J(\hat{t},\hat{\theta})\nonumber\\
&\quad \times\bigl[
    F_{\Delta}(t-t^\prime ,\theta -\theta^\prime)-F_{\Delta}(t^\prime-t ,\theta -\theta^\prime)
\bigr]\bigl[
    F_{\Delta}(\hat{t}-t^\prime ,\hat{\theta} -\theta^\prime)-F_{\Delta}(t^\prime-\hat{t} ,\hat{\theta} -\theta^\prime)\bigr],\\
    F_{\Delta}(t,\theta) &:= \left[-2i\sin\left(\frac{t + \theta - i\epsilon}{2}\right)\right]^{-\Delta}\left[-2i\sin\left(\frac{t - \theta - i\epsilon}{2}\right)\right]^{-\Delta}.
\end{align}
\end{widetext}
Here, we have introduced the $i\epsilon$-prescription as in \cite{Skenderis:2008dg}.
While $S_\mathrm{uni}$ is the standard unitary term, $S_\mathrm{diss}$ reflects the dissipation.

The above generating functional yields multi-point functions \eqref{eq: correlators} by differentiating it with $J$ to set $J=0$.
First, since $Z[J]$ is quadratic in $J$, the one-point function trivially vanishes: $\braket{O(t,\theta)} = 0$.
The two-point function is, for $0<t_1<t_2$, given by
\begin{widetext}
\begin{align}\label{eq: two point}
&\braket{O(t_2,\theta_2)O(t_1,\theta_1)} =
\frac{2(\Delta-1)^2}{\pi}  F_{\Delta}(t_2-t_1,\theta_2-\theta_1)
-
\frac{4\gamma(\Delta-1)^4}{\pi^2}\int \d t \d \theta \Bigl\{\Theta(t_1-t)\Bigr.\nonumber\\
&\qquad\qquad
\Bigl.
\times  \left[
F_{\Delta}(t_1-t ,\theta_1 -\theta)
-F_{\Delta}(t -t_1 ,\theta_1 -\theta)\right]
\left[
F_{\Delta}(t_2-t,\theta_2 -\theta)
-F_{\Delta}(t -t_2 ,\theta_2 -\theta)\right]
\Bigr\}.
\end{align}
\end{widetext}
Again, the first term is the standard unitary two-point function, and the second term is the new contribution.
For general $\Delta$, the UV regulator $\epsilon$ cannot be sent to $0$ \cite{supp}.

\ssection{Comparison with generalized free CFT}
Then, what CFT does the above bulk model describe?
\red{The model, under the probe approximation, is known to be dual to a generalized free CFT$_2$ when $\gamma = 0$, as is evident from \eqref{eq: S_uni}.
Thus, we expect the model with $\gamma \neq 0$ to predict correlators up to $\mathcal{O}(\gamma^1)$ in a generalized free CFT that evolves from the vacuum state under \eqref{eq: Lindblad}.
This is because (a) AdS$_3$ corresponds to the vacuum state, and (b) the probe approximation remains valid as long as the boundary state stays close to the vacuum.
To see a further match between the bulk and boundary beyond $\mathcal{O}(\gamma^1)$, the backreaction to the metric field must perturbatively be accounted for in \eqref{eq: bulk action}.}

Thus, we calculate the one- and two-point functions to \red{$\mathcal{O}(\gamma^1)$} from \eqref{eq: boundary Z}, expanding it with respect to \red{$\gamma$}.
Then, we find that $\braket{O}$ consists of terms as $\braket{0|O|0}$ and $\braket{0|OOO|0}$.
Since the CFT is now a generalized free theory and $\braket{0|O|0} = 0$, we conclude $\braket{O(t,\theta)}=0$ by the Wick theorem.
Similarly, for the two-point function, we find that terms of $\braket{0|OO|0}$ appear in \red{$\mathcal{O}(\gamma^0)$} and those of $\braket{0|OOOO|0}$ appear in \red{$\mathcal{O}(\gamma^1)$}.
The former corresponds to the first unitary term in \eqref{eq: two point}, and the latter to the second (see \cite{supp} for the detail).

\ssection{\red{Comments on relaxation and entropy}}
In the above simple model, can we extract some information about relaxation from \eqref{eq: two point}?
Indeed, \eqref{eq: Lindblad} has a trivial steady state $\rho \propto 1$ (Gibbs state in the high-temperature limit).
Unfortunately, however, calculations up to $\mathcal{O}(\gamma^1)$ reveal nothing about relaxation, since the probe approximation is only applicable while the spacetime remains very close to the vacuum AdS solution (i.e., while the CFT state remains close to the vacuum), and will break down at late times.
To study late-time dynamics, we need to survey beyond $\mathcal{O}(\gamma^1)$.

In fact, the calculation of the entropy up to $\mathcal{O}(\gamma^1)$ tells that the probe approximation cannot capture relaxation.
To obtain the entropy, one computes the partition function on the replica manifold.
In the bulk, the replica manifold is constructed by setting $J=0$ in Fig.\ \ref{fig: bulk_contour} and connecting the forward segment of the $i$-th replica to the $(i+1)$-th backward segment, where the white noise is independent across replicas.
As a result, one finds that the entropy remains zero at $\mathcal{O}(\gamma^1)$.
Since the CFT does not evolve unitarily, the entropy must change to a nonzero value, implying that we must go beyond $\mathcal{O}(\gamma^1)$ to see information about relaxation.
A detailed derivation both in the bulk and the boundary is provided in \cite{supp}.

In the future, by incorporating backreaction on the metric and studying up to $\mathcal{O}(\gamma^2)$, one may observe signatures about relaxation in the correlation functions and a nontrivial time evolution of the entropy.
Alternatively, it would be interesting to find an exact solution in a two-dimensional model such as JT gravity.

\ssection{Summary and discussions}
In this Letter, we developed a method to compute time-ordered correlation functions in the bulk when a holographic CFT is open and governed by the Lindblad equation.
In the bulk, the openness is represented by the white noise $\lambda$ in the asymptotic boundary condition.
The white noise is sampled not independently among the forward and backward segments, shared by them.
We applied the method to a free real scalar field in the bulk and confirmed the consistency with a generalized free CFT.

While we have focused on a real scalar primary operator, the same technique should be applicable to multi-trace operators \cite{Aharony:2001pa, Witten:2001ua, Berkooz:2002ug, Sever:2002fk, Aharony:2005sh, Aharony:2006hz} and other bosonic operators as well.
In such cases, the methods developed in \cite{gullo2014non} may be useful.
It enables the construction of Lindblad dynamics by simply adding noise to the unitary AdS/CFT setup from the beginning, making it relatively easy to obtain a bulk dual picture.
Moreover, the techniques in \cite{gullo2014non} can also be used to extend the framework to non-Markovian processes by replacing white noise with colored one.

So far, the AdS/CMP program has provided models for a wide range of condensed matter theories, offering predictions for thermodynamic aspects such as phase diagrams as well as for measurable quantities like operator expectation values.
As the present work has established for the first time a general framework to treat the most basic open quantum systems within AdS/CFT, we expect this to pioneer ``open AdS/CMP".
For example, we can consider holographic superconductors \cite{Hartnoll:2008vx, Hartnoll:2008kx} exposed to an environment.

The benefits of our framework are not only for condensed matter but also for quantum gravity.
One key example is black hole (BH) evaporation.
Since BH evaporation involves quantum gravitational effects that are notoriously difficult to describe, much of the progress has relied on studying asymptotically AdS spacetimes, where insights from AdS/CFT can be leveraged.
To prevent the Hawking radiation from reflecting off the AdS boundary, previous studies have attached thermal reservoirs beyond the AdS boundary to allow the radiation to escape \cite{Rocha:2008fe, Almheiri:2019psf, Penington:2019npb, Almheiri:2019qdq}.
Now that we can handle the gravitational picture dual to open quantum systems, our framework may offer a new playground for exploring the black hole evaporation problem.
Furthermore, \cite{Glatthard:2025mbb} pointed out that the time evolution of entropy in certain Lindblad systems resembles the behavior of the Page curve \cite{Page:1993wv} of Hawking radiation.
Also from this perspective, there is reason to expect our method to bring new insights to this problem.

Another application lies in black hole thermodynamics.
There are efforts to reformulate black hole thermodynamics purely as the quantum thermodynamics in the dual CFT \cite{Takeda:2024qbq, Shigemura:2024yeb}.
In the series of work, coarse-graining is introduced via the principle of maximum entropy.
Since Lindblad dynamics inherently includes coarse-graining, the von Neumann entropy of the state serves directly as a meaningful coarse-grained entropy.
In fact, for certain choices of jump operators, a Lindblad system relaxes toward the Gibbs state, and the familiar form of the second law can be derived (for a review, \cite{e15062100}).
This suggests a new direction in which one can study black hole relaxation processes and the second law, from the perspective of open systems in AdS/CFT.

\ssection{Acknowledgments}
We would like to thank Ryusuke Hamazaki for the intensive lecture in Kyoto University.
We are also grateful to Koji Hashimoto and Sotaro Sugishita for comments on our idea.
D.T.\ was supported by Grant-in-Aid for JSPS Fellows No.\ 22KJ1944 and is also by RIKEN Special Postdoctoral Researchers Program.

\bibliography{ref}


\clearpage
\onecolumngrid
\appendix
\setcounter{page}{1}

\begin{center}
\fontsize{14pt}{12pt}{\textbf{Supplemental Material}}
\end{center}

\section{Derivation of \eqref{eq: bulk Z} from \eqref{eq: boundary Z}}
Fist, introducing auxiliary fields $\lambda_{f,b}$ and $\eta_{f,b}$, \eqref{eq: boundary Z} can be deformed as
\begin{align}
Z[J] = &\int\mathcal{D}\phi\mathcal{D}\lambda\mathcal{D}\eta\; e^{iI[C;\phi]}\exp\Bigl[\int\d^{d}x\,\Bigl(
i (J(x)+\lambda_f(x)) O_f(x) -i\lambda_b(x)O_b(x)
\nonumber\\
&\qquad\qquad\qquad
+ \gamma \eta_f(x) \eta_b(x)- \frac{\gamma }{2}\eta_f(x)^2 - \frac{\gamma }{2}\eta_b(x)^2 - i\lambda_f(x)\eta_f(x) + i\lambda_b(x)\eta_b(x)
\Bigr) 
\Bigr].
\end{align}
In fact, performing $\mathcal{D}\lambda$ and $\mathcal{D}\eta$ yields \eqref{eq: boundary Z}.
Since we can now use the standard AdS/CFT dictionary, $Z[J]$ can be written as
\begin{align}
Z[J] = &\int \mathbb{D}\Phi \mathcal{D}\lambda\mathcal{D}\eta\; e^{i S[M;\Phi;J+\lambda_f,\lambda_b]}\nonumber\\
&\qquad
\times
\exp\left[\int \d^d x\; \left(
\gamma \eta_f(x) \eta_b(x)- \frac{\gamma }{2}\eta_f(x)^2 - \frac{\gamma }{2}\eta_b(x)^2 - i\lambda_f(x)\eta_f(x) + i\lambda_b(x)\eta_b(x)
\right)\right].
\end{align}
Then, noting that
\begin{align}
    &\gamma \eta_f(x) \eta_b(x)- \frac{\gamma }{2}\eta_f(x)^2 - \frac{\gamma }{2}\eta_b(x)^2 - i\lambda_f(x)\eta_f(x) + i\lambda_b(x)\eta_b(x)\nonumber\\
    &=
    -\frac{\gamma}{2}\left(\eta_f - \eta_b + i\frac{\lambda_f}{\gamma} \right)^2 - \frac{\lambda_f^2}{2\gamma} + i \eta_b (\lambda_b - \lambda_f),
\end{align}
we finally obtain
\begin{align}
    Z[J] = &\int \mathbb{D}\Phi \mathcal{D}\lambda_f\; e^{i S[M;\Phi;J+\lambda_f,\lambda_f]}
\exp\left[-\frac{1}{2\gamma}\int \d^d x\; \lambda_f(x)^2
\right],
\end{align}
which is nothing but \eqref{eq: bulk Z}.

\section{Derivation of \eqref{eq: example Z}}
We compute the solution of $(\nabla^2-m^2)\Phi=0$ with the boundary condition \eqref{eq: BC}.
Below, we call the Euclidean segment attached to the forward (backward) segment as $\mathrm{E}f$ ($\mathrm{E}b$).
With a slight generalization of 4.1.2.\ in \cite{Skenderis:2008dg}, the solution is given as follows\footnote{
As in \cite{Skenderis:2008dg}, we first introduce a large time cutoff $T$ for the Lorentzian segments, and after the computation we take $T\to \infty$.
}:
\begin{align}
&\Phi_f(t,r,\theta) = \int_{\rm F} \d\omega  \sum_{k} e^{-i \omega t+ik\theta} \chi_f(\omega,k)\tilde{\Phi}(r,\omega,|k|) - \sum_{n,k} \chi_b(-\omega_{nk},k)e^{i \omega_{nk}t +ik\theta}g(r, \omega_{nk}, k),\label{eq: sol_f}
\\
&\Phi_b(t,r,\theta) = \int_{{\rm F}} \d\omega  \sum_{k} e^{-i \omega t+ik\theta} \chi_b(\omega,k)\tilde{\Phi}(r,\omega,|k|) - \sum_{n,k} \chi_f(\omega_{nk},k)e^{-i \omega_{nk}t_3 +ik\theta}g(r, \omega_{nk}, k),\label{eq: sol_b}
\\
&\Phi_{\mathrm{E}f}(\tau,r,\theta) = \sum_{n,k} [\chi_f(-\omega_{nk},k)- \chi_b(-\omega_{nk},k)]e^{\omega_{nk}\tau+ik\theta}g(r, \omega_{nk}, k),
\\
&\Phi_{\mathrm{E}b}(\tau,r,\theta) = -\sum_{n,k} [\chi_f(\omega_{nk},k)- \chi_b(\omega_{nk},k)]e^{-\omega_{nk}\tau+ik\theta}g(r, \omega_{nk}, k),
\\
&\tilde{\Phi}(r,\omega,|k|) = r^{|k|}(1+r^2)^{-\frac{|k|+\Delta}{2}} C(\omega,|k|)\; _2F_1\left[\frac{|k|+\Delta+\omega}{2},\frac{|k|+\Delta-\omega}{2},1+|k|;\frac{r^2}{r^2+1}\right].
\label{eq: Phi tilde}
\end{align}
Here are definitions in order.
First we have defined $\omega_{nk}:=(2n+|k|+\Delta)$, $\chi_f(\omega,k):=J(\omega,k)+\lambda(\omega,k)$, and $\chi_b(\omega,k):=\lambda(\omega,k)$, where $J(\omega,k)$ and $\lambda(\omega,k)$ are the Fourier components of $J(t,\theta)$ and $\lambda(t,\theta)$, respectively.
The $\omega$-contour $\rm F$ indicates the prescription to avoid poles on the real axis, and here we chose it as the Feynman contour; another convention just requires a modification to the coefficients of $g(r, \omega_{nk}, k)$.
The coefficient $C(\omega,|k|)$ is chosen so that $\tilde{\Phi} \sim 1\times r^{\Delta-2}$ as $r\to \infty$:
\begin{align}
    C(\omega, |k|) = \frac{\Gamma(\frac{|k|+\Delta+\omega}{2})\Gamma(\frac{|k|+\Delta-\omega}{2})}{\Gamma(\Delta-1)\Gamma(1+|k|)}.
\end{align}
In the expression of $\tilde{\Phi}$, $_2F_1$ represents the hypergeometric function.
Finally, $g(r, \omega_{nk}, k)$ is the normal mode and defined as
\begin{align}
   g(r, \omega_{nk}, k) = \oint_{\omega_{nk}}\d\omega\;\tilde{\Phi}(r,\omega,|k|) = \frac{4\pi i}{n!} \frac{\Gamma(n+\Delta) \Gamma(n+\Delta+|k|)}{\Gamma(\Delta-1) \Gamma(\Delta)\Gamma(n+|k|+1)} r^{-\Delta}(1 + \cdots).
\end{align} 

To compute the on-shell action, we do not need the Euclidean solution, owing to \eqref{eq: Euclidean BC}.
From \eqref{eq: sol_f}, \eqref{eq: sol_b} and the counterterms in \cite{deHaro:2000vlm}, the on-shell action becomes 
\begin{align}
S=& 
\frac{2(\Delta -1)}{\pi\Gamma(\Delta-1)\Gamma(\Delta)}\int\d t \d\theta\d\hat{t}\d\hat{\theta}\sum_{n,k}e^{ik(\theta-\hat{\theta})}\frac{\Gamma(n+\Delta)\Gamma(n+\Delta+|k|)}{\Gamma(n+1)\Gamma(n+1+|k|)}\Theta(t-\hat{t})\nonumber\\
&\qquad\qquad\qquad\qquad\qquad\qquad\qquad\qquad\times\left[
iJ(t,\theta)J(\hat{t},\hat{\theta}) e^{i\omega_{nk}(t-\hat{t})}
-2J(t,\theta)\lambda(\hat{t},\hat{\theta}) \sin (\omega_{nk}(t-\hat{t}))
\right].\label{eq: on-shell S}
\end{align}

Next, we perform the $\lambda$-integral of \eqref{eq: bulk Z}.
Since \eqref{eq: on-shell S} is linear in $\lambda$,  \eqref{eq: bulk Z} is just a Gaussian integral.
Then, the integral results in $Z[J]=e^{W[J]}$ with $W[J]$ given as
\begin{align}
&W[J]=-\frac{2(\Delta -1)}{\pi\Gamma(\Delta-1)\Gamma(\Delta)}\int\d t\d\theta\d\hat{t}\d\hat{\theta}
\sum_{n,k}\frac{\Gamma(n+\Delta)\Gamma(n+\Delta+|k|)}{\Gamma(n+1)\Gamma(n+1+|k|)}e^{ik(\theta-\hat{\theta})+i\omega_{nk}(t-\hat{t})}\Theta(t-\hat{t})J(t,\theta)J(\hat{t},\hat{\theta})
\notag\\
-2 &\gamma \int\d t\d\theta
\Bigl[
\frac{2(\Delta -1)}{\pi\Gamma(\Delta-1)\Gamma(\Delta)}
\int\d t^\prime\d\theta^\prime\sum_{n,k}\frac{\Gamma(n+\Delta)\Gamma(n+\Delta+|k|)}{\Gamma(n+1)\Gamma(n+1+|k|)}e^{ik(\theta^\prime-\theta)}\sin(\omega_{nk}(t^\prime-t))
\Theta(t^\prime-t)J(t^\prime,\theta^\prime)
\Bigr]^2.
\end{align}
Note that the overall constant of $Z[J]$ has been neglected.
With the introduction of $i \epsilon$ factor, the summation that appears above converges as follows:
\begin{align}
\sum_{n,k}\frac{\Gamma(n+\Delta)\Gamma(n+\Delta+|k|)}{\Gamma(n+1)\Gamma(n+1+|k|)}e^{i\omega_{nk}(t - i\epsilon)+ik\theta}=
\Gamma(\Delta)^2e^{i\Delta (t- i\epsilon)}( 1-e^{i(t- i\epsilon+\theta)} )^{-\Delta}( 1-e^{i(t- i\epsilon-\theta)} )^{-\Delta}.
\end{align}
Thus, we conclude \eqref{eq: example Z}.

\section{CFT computation \red{of two-point function}}
Here, we detail the CFT computation of the two-point function.
We start with \eqref{eq: boundary Z} with $d=2$ and assume that the CFT is a generalized free theory when $\gamma =0$, meaning that the Wick theorem applies to vacuum correlators.

First, the two-point function is, from \eqref{eq: boundary Z}, given by
\begin{align}
    \braket{O(t_2,\theta_2)O(t_1,\theta_1)}
    =
    \frac{1}{Z[0]}
   \int\mathcal{D}\phi\, e^{iI[C;\phi]} \exp\left[\gamma\int_{t=0}^{t=\infty} \d^{2}x\,\left( O_f O_b- \frac{1 }{2}O_f^2 - \frac{1 }{2}O_b^2\right)\right]O_f(t_2,\theta_2)O_f(t_1,\theta_1).
\end{align}
Below, we set $x_1 = (t_1,\theta_1)$ and $x_2 = (t_2,\theta_2)$.
As explained in Letter, we are interested in terms to $\mathcal{O}(\gamma^1)$.
Recalling that the Euclidean segments just create vacuum state, we obtain \red{$\mathcal{O}(\gamma^0)$ and $\mathcal{O}(\gamma^1)$} terms as follows: for $0<t_1<t_2$,
\begin{align}
   \braket{O(x_2)O(x_1)}_0 :=& \braket{0|O(x_2)O(x_1)|0} = c\, F_\Delta(t_2-t_1, \theta_2-\theta_1),\\
   \braket{O(x_2)O(x_1)}_1
   :=&\;
   \frac{\gamma}{Z[0]}\int\mathcal{D}\phi\, e^{iI[C;\phi]} \left[\int_{t=0}^{t=\infty} \d^{2}x\,\left(O_f O_b- \frac{1}{2}O_f^2 - \frac{1}{2}O_b^2\right)\right]O_f(x_2)O_f(x_1)\\
   =&\; \gamma\int_{t=0}^{t=\infty} \d^2x\; \Bigl[\braket{0|O(x)\mathrm{T}[O(x)O(x_2)O(x_1)]|0}\nonumber\\
   &\qquad\qquad -\frac{1}{2}\braket{0|\mathrm{T}[O(x)^2O(x_2)O(x_1)]|0} - \frac{1}{2}\braket{0|O(x)^2O(x_2)O(x_1)|0} \Bigr],\label{eq: <OO>_2}
\end{align}
where $c$ is the normalization constant and $\mathrm{T}$ is the time-ordering symbol.
Note that the forward operators, including $O(x_{1,2})$, are put on the right in the time-ordering, while the backward operators are put on the left.

Let us calculate $\braket{O(x_2)O(x_1)}_1$ furthermore.
We find that there is no contribution from $t_1<t$ in the integral of \eqref{eq: <OO>_2}, as follows.
First, for $t_2<t$, it is obvious that the three terms cancel out with each other.
Second, for $t_1<t<t_2$, the integrand becomes
\begin{align}
    \braket{0|O(x)O(x_2)O(x)O(x_1)|0} -\frac{1}{2}\braket{0|O(x_2)O(x)^2O(x_1)|0} - \frac{1}{2}\braket{0|O(x)^2O(x_2)O(x_1)|0}.
\end{align}
Using the Wick theorem, the four-point correlators are decomposed into products of two-point functions.
For example, we have
\begin{align}
    &\braket{0|O(x)O(x_2)O(x)O(x_1)|0}\nonumber \\
    &= \braket{0|O(x)O(x_2)|0}\braket{0|O(x)O(x_1)|0} + \braket{0|O(x)^2|0}\braket{0|O(x_2)O(x_1)|0} + \braket{0|O(x_2)O(x)|0}\braket{0|O(x)O(x_1)|0}.
\end{align}
Repeating the similar process also for the remaining four-point functions, we see that all the terms cancel out.

\begin{figure*}[t]
\centering
    \begin{minipage}{0.45\textwidth}
        \centering
        \includegraphics[width=0.9\linewidth]{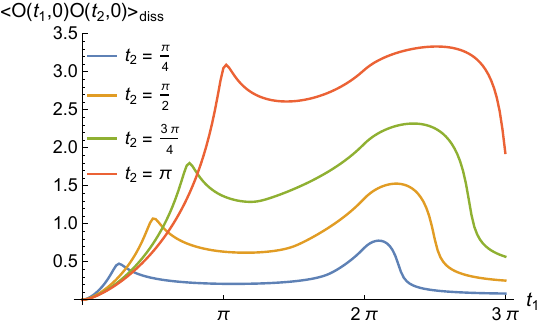}
    \end{minipage}
    \hspace{20pt}
    \begin{minipage}{0.45\textwidth}
        \centering
        \includegraphics[width=0.9\linewidth]{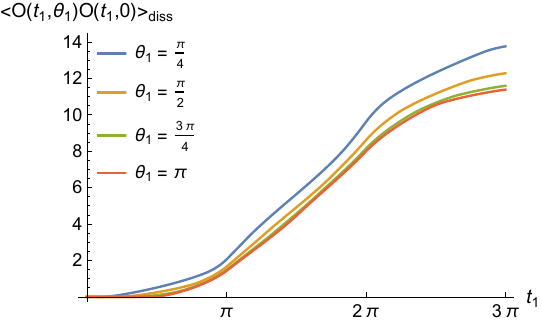}
    \end{minipage}
    \caption{The plots show the dissipation part of the time-ordered two-point function for $\Delta = 3/10$ and $\epsilon = 1/20$.
    Left: $\theta_1 = \theta_2 = 0$. Note that for $t_2 < t_1$, the roles of $t_1$ and $t_2$ are exchanged in \eqref{eq: two point}.
    Right: $t_2 = t_1$ and $\theta_2 = 0$.
    }
    \label{fig: two-point}
\end{figure*}

Thus, we now have
\begin{align}
    \braket{O(x_2)O(x_1)}_1
   =&\; \gamma\int_{t=0}^{t=t_1} \d^2x\; \Bigl[\braket{0|O(x)O(x_2)O(x_1)O(x)|0}\nonumber\\
   &\qquad\qquad -\frac{1}{2}\braket{0|O(x_2)O(x_1)O(x)^2|0} - \frac{1}{2}\braket{0|O(x)^2O(x_2)O(x_1)|0} \Bigr].
\end{align}
Again the Wick theorem reduces it to
\begin{align}
    \braket{O(x_2)O(x_1)}_1 &= -\gamma\int_{t=0}^{t=t_1} \d^2x\; \braket{0|[O(x_1),O(x)]|0}\braket{0|[O(x_2),O(x)]|0}\nonumber\\
    &= -\gamma c^2\int_{t=0}^{t=t_1} \d^2x\;
    \left[
    F_{\Delta}(t_1-t ,\theta_1 -\theta)
    -F_{\Delta}(t -t_1 ,\theta_1 -\theta)\right]\nonumber\\
    &\hspace{120pt}
    \times
    \left[
    F_{\Delta}(t_2-t,\theta_2 -\theta)
    -F_{\Delta}(t -t_2 ,\theta_2 -\theta)\right].
\end{align}
Setting $c = 2(\Delta -1 )^2/\pi$, we now find that $\braket{O(x_2)O(x_1)}_0 + \braket{O(x_2)O(x_1)}_1$ is equal to \eqref{eq: two point}.
\red{FIG.\ \ref{fig: two-point} shows plots of the two-point function. 
}

\section{\red{The entropy vanishes at $\mathcal{O}(\gamma^1)$}}
In this section, we calculate the entropy up to $\mathcal{O}(\gamma^1)$, both in the bulk and the boundary.
As a result, we find that the entropy vanishes even at $\mathcal{O}(\gamma^1)$.
Thus, further calculations at $\mathcal{O}(\gamma^2)$ is necessary to see the non-trivial leading term.
(The entropy must evolve in time, since the state will not be kept pure.)

\begin{figure}
    \centering
    \includegraphics[width=0.7\linewidth]{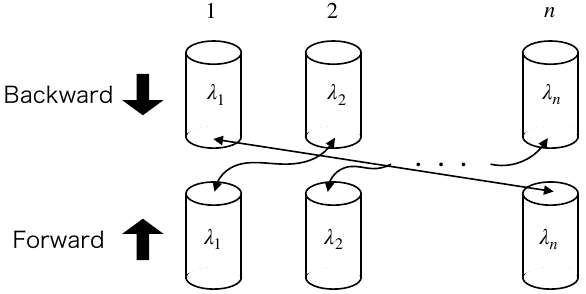}
    \caption{The replica manifold $M_n$ to compute the entropy. While the Euclidean segments, which create the initial vacuum state, are omitted in the figure, we note that they are attached to each Lorentzian segment here as well as in FIG.\ \ref{fig: bulk_contour}.}
    \label{fig: replica}
\end{figure}

\subsection{Bulk calculation}
In the gravity side, we evaluate the on-shell action on the replica manifold $M_n$, which is drawn in FIG.\ \ref{fig: replica}.
The white noise is introduced independently to each replica, since each $\rho$ consisting in $\mathrm{Tr}\rho^n$ has one white noise variable.
The action is evaluated as\footnote{
If we were to take the all the white noise variables identical, the action would trivially vanish.
}
\begin{align}
iS[M_n;\Phi;\lambda_1,\cdots,\lambda_n] =-\frac{(\Delta-1)^2}{\pi}\int_{t,t' \in [0,T]} \d^2x\d^2x^\prime \vec{\lambda}^\top X\ \vec{\lambda},
\end{align}
where $T$ is the time when we want to measure the entropy, $ \vec{\lambda}=(\lambda_1,\cdots,\lambda_n)^\top$, and $X$ is an $n\times n$ matrix given by
\begin{equation*}
   \begin{cases}
      X_{ii} = F_{\Delta}(t-t^\prime,\theta-\theta^\prime)+F_{\Delta}(t'-t,\theta'-\theta) \\
      X_{i\hspace{1pt}i+1} =X_{i+1\hspace{1pt}i}= -\frac{1}{2}[F_{\Delta}(t-t^\prime,\theta-\theta^\prime)+F_{\Delta}(t'-t,\theta'-\theta)]\\
      X_{ij} = 0\qquad (\mbox{otherwise})
   \end{cases}.
\end{equation*}
We have adopted the convention that $i=n+1$ represents $i=1$.
In deriving this, \eqref{eq: sol_f} -- \eqref{eq: Phi tilde} are reusable.

Therefore, the partition function $Z_n$ is computed as
\begin{align}
Z_{n} &= \int \mathcal{D}\lambda e^{iS[M_n;\Phi;\lambda_1,\cdots,\lambda_n]-\frac{1}{2\gamma}\int \d^2x\d^2x^\prime \vec{\lambda}^\top {\rm I}_{n} \delta(t-t^\prime) \delta(\theta-\theta^\prime) \vec{\lambda}}\nonumber\\
&\propto \left(\det\left[{\rm I}_n \delta + 2\gamma \frac{(\Delta-1)^2}{\pi}X\right]\right)^{-1/2}\nonumber\\
&\simeq 1 - \gamma \frac{(\Delta-1)^2}{\pi}\mathrm{Tr}X\nonumber\\
&=1 -  4n\gamma T(\Delta-1)^2 F_\Delta(0,0)
\end{align}
In the second line, we have omitted irrelevant factors including factors like $(\mathrm{const})^n$, and in the third line, we have extracted the terms up to $\mathcal{O}(\gamma)$.
The determinant and the trace run also over the function space.
From $Z_n$, the holographic entropy $\mathcal{S}(T)$ is given by
\begin{align}
 \mathcal{S}(T) = -\left(\frac{\partial}{\partial n} \left. \frac{1}{n} \ln Z_n\right)\right|_{n\to 1} = o(\gamma).
\end{align}

\subsection{Boundary calculation}
In the CFT side, the partition function $Z_n$ can be deformed up to $\mathcal{O}(\gamma^1)$ as
\begin{align}
Z_n &= \int \mathcal{D} \phi e^{iI[\partial M_n;\phi]   } \exp\Bigl[-\frac{\gamma}{2}\int_{t\in [0,T]} \d^2 x(O_{ f1}(x)^2 + O_{ b1}(x)^2 +\cdots +O_{{ f}n}(x)^2 + O_{{ b}n}(x)^2 ) \nonumber\\
&\hspace{150pt}+\gamma \int_{t\in [0,T]} \d^2x( O_{ f1}(x) O_{ b1}(x) +\cdots +O_{fn}(x) O_{ bn}(x) )\Bigr]
\nonumber\\
&\simeq \int \mathcal{D}\phi e^{iI[\partial M_n;\phi]}\Bigl[
1 -\frac{\gamma}{2}\int \d^2 x(O_{ f1}(x)^2 + O_{ b1}(x)^2 +\cdots +O_{{ f}n}(x)^2 + O_{{ b}n}(x)^2 ) \nonumber\\
&\hspace{150pt}+\gamma \int \d^2x( O_{ f1}(x) O_{ b1}(x) +\cdots +O_{{ f}n}(x) O_{ bn}(x) ) \Bigr]\nonumber\\
&\propto 1 - 2n\times\frac{\gamma}{2} \int \d^2x \braket{0|O(x)^2|0} + n\times \gamma \int \d^2x\braket{0|O(x)|0}^2 \nonumber\\
&= 1 - n \gamma c \int \d^2x\ F_{\Delta}(0,0) \nonumber\\
&=1 - 2\pi n\gamma (\Delta-1)^2 cF_{\Delta}(0,0) .
\end{align}
Here, $O_{{ f}i}$ ($O_{{ b}i}$) is the operator inserted in the forward (backward) segment of the $i$-th replica.
Since we have identified $c=2(\Delta-1)^2/\pi$, it matches the partition function of the bulk. As a result, the entropy again vanishes at $\mathcal{O}(\gamma^1)$.
Note that since we did not use the large $N$ factorization, this result does not depend on the details of the bulk model.

\section{On the divergence of \eqref{eq: two point}}
\begin{figure}[t]
    \centering
    \includegraphics[height = 6cm]{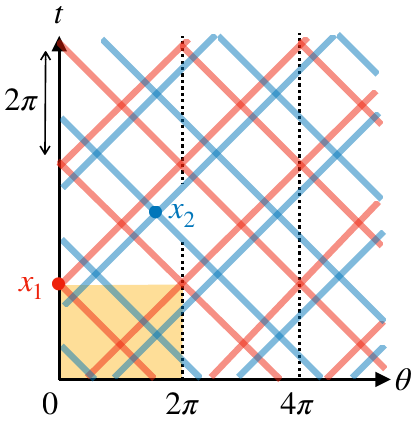}
    \caption{
    The light-cone of $x_1 = (t_1, \theta_1)$ (red lines) and that of $x_2 = (t_2, \theta_2)$ (blue lines), and the integration domain of $x = (t, \theta)$ (orange shaded region) in \eqref{eq: two point}.
    The light-cones repeat with period $2\pi$ due to $\theta \sim \theta + 2\pi$.
    }
    \label{fig: light-cones}
\end{figure}

Here, we use the notation $x=(t,\theta)$.
The two-point function \eqref{eq: two point} diverges for general $\Delta$ if we take $\epsilon\to 0$.
As is well-known, the unitary part $F(t_2 - t_1,\theta_2 - \theta_1)$ diverges only when there exists a null geodesic connecting $x_1$ and $x_2$, i.e., when $x_2$ is on the red lines in FIG.\ \ref{fig: light-cones}.
Here, we consider the possible divergence from the dissipation part, which is the second term of \eqref{eq: two point} and we call $G(x_1,x_2)$ hereafter.

The integrand of $G(x_1,x_2)$ diverges when there exist one or more null geodesics that connect $x$ to $x_1$ or $x_2$.\footnote{
Note that $F_{\Delta}(t_1-t ,\theta_1 -\theta)$ and $F_{\Delta}(t -t_1 ,\theta_1 -\theta)$ do not cancel out the individual divergence.
}
FIG.\ \ref{fig: light-cones} helps us know when $x$ is on the light-cone of $x_1$ and that of $x_2$.
The strength of the divergence of the integrand depends on how many lines pass through $x$, regardless of the colors.
If there are $n$ ($n = 0,1,2,3,4$) lines that pass through $x$, then the integrand behaves as $\mathcal{O}(\epsilon^{-n \Delta})$ at the point.
On the other hand, if we evaluate the integration of $G(x_1,x_2)$ by introducing a lattice cut-off $\epsilon$, the integral measure $\d t \d \theta$ can be seen as $\mathcal{O}(\epsilon^{2})$.

First, let us consider the case shown in FIG.\ \ref{fig: light-cones}, where $x_2$ does not lie on any null line emanating from $x_1$.
In the shaded region, there are infinitely many $n=1$ points along the null lines, as well as a finite number of $n=2$ points.
The lattice regulation implies that the order of the number of $n=1$ points is $\mathcal{O}(\epsilon^1)$.
Thus, $G(x_1,x_2)$ includes contributions of $\mathcal{O}(\epsilon^{1-\Delta})$ and $\mathcal{O}(\epsilon^{2-2\Delta})$, and hence $G(x_1,x_2)$ will diverge when $\Delta\ge 1$.

Similarly, we can also evaluate the divergence when there exists one or more null geodesics that pass through $x_2$.
In either case, we see that $G(x_1,x_2)$ diverges for $\Delta \ge 1/2$.

\end{document}